\documentclass[Physsubmission, Phys]{SciPost}

\usepackage{graphicx}
\usepackage{caption}
\usepackage{subcaption}
\hypersetup{
    colorlinks,
    linkcolor={red!50!black},
    citecolor={blue!50!black},
    urlcolor={blue!80!black}
}

\usepackage[bitstream-charter]{mathdesign}
\newcommand{\numberUnit}[2]{$#1\mbox{ #2}$}

\DeclareSymbolFont{usualmathcal}{OMS}{cmsy}{m}{n}
\DeclareSymbolFontAlphabet{\mathcal}{usualmathcal}
\begin{document}

\begin{center}{\Large \textbf{
Results on high energy
galactic cosmic rays from
the DAMPE space mission}}\end{center}

% TODO: write the author list here. Use initials + surname format.
% Separate subsequent authors by a comma, omit comma at the end of the list.
% Mark the corresponding author with a superscript *.
\begin{center}
Leandro Silveri\textsuperscript{1,2 $\star$}\\\vspace*{0.5cm}
(on behalf of the DAMPE Collaboration)\\
\end{center}

\begin{center}
{\bf 1} Gran Sasso Science Institute (GSSI), Via Iacobucci 2, I-67100 L’Aquila, Italy
\\
{\bf 2} Istituto Nazionale di Fisica Nucleare (INFN) - Laboratori Nazionali del Gran Sasso, Italy
\\
* leandro.silveri@gssi.it
\end{center}

\begin{center}
\today
\end{center}

% For convenience during refereeing (optional),
% you can turn on line numbers by uncommenting the next line:
%\linenumbers

\definecolor{palegray}{gray}{0.95}
\begin{center}
\colorbox{palegray}{
  \begin{tabular}{rr}
  \begin{minipage}{0.1\textwidth}
    \includegraphics[width=30mm]{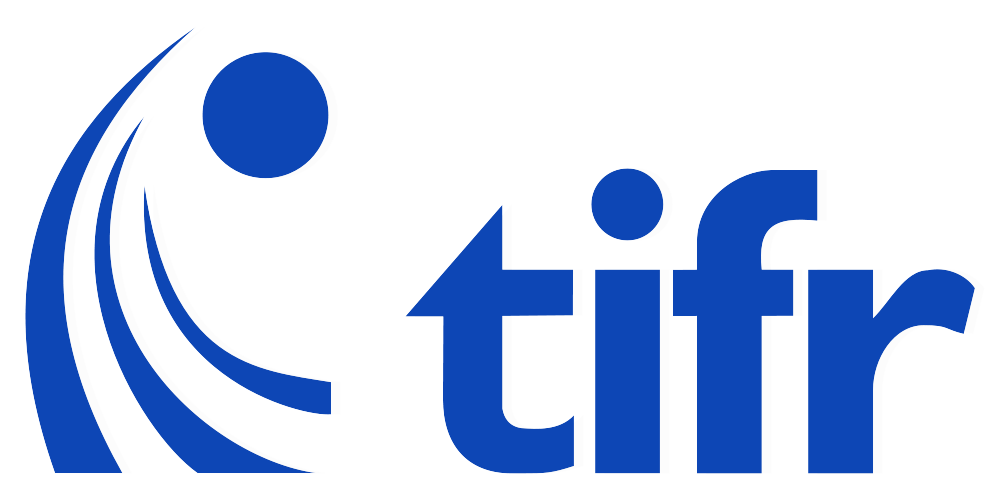}
  \end{minipage}
  &
  \begin{minipage}{0.85\textwidth}
    \begin{center}
    {\it 21st International Symposium on Very High Energy Cosmic Ray Interactions (ISVHE- CRI 2022)}\\
    {\it Online, 23-27 May 2022} \\
    \doi{10.21468/SciPostPhysProc.202205002}\\
    \end{center}
  \end{minipage}
\end{tabular}
}
\end{center}

\section*{Abstract}
{\bf
% TODO: write your abstract here.
%The abstract is in boldface, and should fit in 8 lines.
%It should be written in a clear and accessible style, emphasizing the context, the problem(s) studied, the methods used, the results obtained, the conclusions reached, and the outlook. You can add a table contents, recommended if your paper is more than 6 pages long.
DAMPE (Dark Matter Particle Explorer) is a satellite-born experiment launched in 2015 in a sun-synchronous 
orbit at 500 km altitude, and it has been taking data in stable conditions ever since. Its main goals
include the spectral measurements up to very high energies, cosmic electrons/positrons and gamma rays
up to tens of TeV, and protons and nuclei up to hundreds of TeV.
The detector's main features include the 32 radiation lengths deep calorimeter and large geometric
acceptance, making DAMPE one of the most powerful space instruments in operation, covering with
high statistics and small systematics the high energy frontier up to several hundreds TeV.
The results of spectral measurements of different species are shown and discussed.
}

\vspace{10pt}
\noindent\rule{\textwidth}{1pt}
\tableofcontents\thispagestyle{fancy}
\noindent\rule{\textwidth}{1pt}
\vspace{10pt}

\section{Introduction}
\label{sec:intro}
% TODO: write your article here.
%The unprecedented improvement on both precision and energy sensitivity on Cosmic Rays direct measurements, made available by satellite-borne 
%missions in the last decade,
%opened for the achievement of unexpected results, 
Satellite-borne missions in the last decade have achieved unprecedented improvements both in precision and energy sensitivity for direct measurements of cosmic rays. This has unveiled unexpected results,
such as the observation of a break from single power-law behaviour even prior to the knee region
in multiple nuclei spectra. Among the experiments currently in orbit, DAMPE\cite{DAMPE_mission} is the one capable of reaching the hundreds of TeV energy 
with good statistics because of its 
high acceptance and the deep calorimeter. This makes it possible to compare the results with some ground-based CR detection facilities.
The acceptance for electrons is $\approx 0.3\mbox{ m}^2\mbox{ sr}$ for energies above $\approx 10 \mbox{ GeV}$.

\subsection{DAMPE Space Mission}
\label{sec:dampe}
DAMPE was launched on December 17, 2015, from the Jiuquan Satellite Launch Center, China, in a sun-synchronous orbit at 500 km altitude.
The mission goals are:
\begin{itemize}
	\item Very accurate measurements of cosmic-ray spectra of electrons, protons and nuclei;
	\item Perform $\gamma$-ray astronomy;
	\item Measure $\gamma$ spectral lines that could point to possible Dark Matter self-interaction channels.
\end{itemize} 

\subsection{DAMPE Sub-Detector modules}
\begin{figure}[h]
\centering
\includegraphics[width=0.5\textwidth]{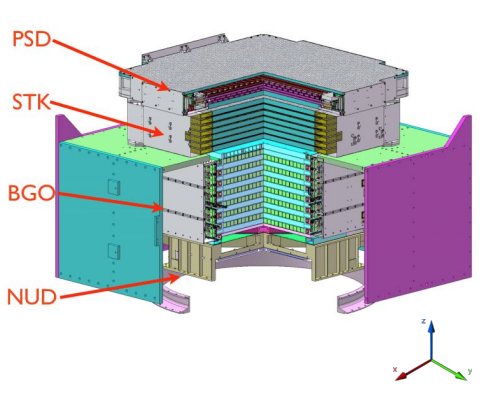}
\caption{DAMPE with its sub-detector modules, from top to bottom: PSD (Plastic Scintillator Detector), 
	STK (Silicon-Tungsten tracKer), BGO (BGO Imaging Calorimeter) and NUD (NeUtron Detector) }
\label{fig:dampe_detector}
\end{figure}

The DAMPE detector is made out of 4 sub-detectors (fig. \ref{fig:dampe_detector}), each one having a different task in order to successfully 
discriminate the nature of the impinging particles, track their direction and measure their energy.

More specifically, starting from the top and going toward the bottom of the detector like shown in fig. \ref{fig:dampe_detector}:
\begin{itemize}
	\item Plastic Scintillator Detector: it is made by 4 layers of plastic scintillator bars used to perform charge measurement and gamma anticoincidence, 
		two on each view (X and Y) in order to reconstruct the information of the impact point, and staggered to improve hermeticity;
		
	\item Silicon-Tungsten Tracker: it consists of 6 planes of silicon microstrip detectors, each one featuring an X and Y segmented layer, with some of them 
		interleaved with tungsten plates in order to enhance the gamma conversion probability for gamma tracking. This results in an angular resolution at normal
		incidence at $100\mbox{ GeV}$ of $\approx 0.1^o$;
		
	\item BGO Imaging Calorimeter: it embodies 14 layers of BGO bars, and the bars are positioned inside a single layer
		along the X or the Y direction of the detector reference frame, alternating them in such a way that allows for the possibility
		of reconstructing the image of the shower, as well as performing the energy measurement. The obtained shower image makes it possible to distinguish
		hadronic from electromagnetic showers. The total depth of this calorimeter is $32\mbox{ X}_0$ (radiation lengths) and $1.6\mbox{ }\Lambda_I$ 
		(interaction lengths) ;
	
	\item Neutron Detector: it is composed of 4 tiles of boron-loaded plastic scintillator placed in a single layer, allowing for the 
		possibility to detect neutrons, providing a further discrimination power against electromagnetic showers.
		
\end{itemize}

A more detailed description is given in \cite{DAMPE_mission}.

\vspace*{\stretch{1}}
\newpage

\section{Proton and Helium Results}
\label{sec:pHe}
Using the data collected by DAMPE from January 2016 to June 2018 (\numberUnit{30}{months}), it was possible to perform the measurement of the proton spectrum 
from \numberUnit{40}{GeV} to \numberUnit{100}{TeV}\cite{DAMPE_protons}. 
The result confirmed a hardening which was already observed by other experiments at 
\numberUnit{500}{GeV}\cite{AMS_protons}\cite{PAMELA_protons}\cite{ATIC_protons}\cite{CREAM_protons}\cite{NUCLEON_protons}.
%and heavily hints to a softening at \numberUnit{14}{TeV} 
Measurements also show a softening at \numberUnit{14}{TeV}
with a strong evidence of $4.7\mbox{ }\sigma$ significance. 
These breaks pointed out that a broken power law
model describes data in a more appropriate way when compared to the single power law spectrum.

%\subsection{Helium}
A similar result was also obtained for helium: 
the DAMPE He spectrum was obtained using 54 months of data (from January 2016 to June 2020), and it has been published for the energy range
starting from \numberUnit{70}{GeV} up to \numberUnit{80}{TeV} of total kinetic energy\cite{DAMPE_helium}. 
It confirmed a hardening which was previously observed by other 
experiments\cite{PAMELA_protons}\cite{ATIC_protons}\cite{CREAM_protons}\cite{AMS_helium}\cite{NUCLEON_helium} and showed a very strong
evidence for a softening at \numberUnit{34}{TeV} with $4.3\mbox{ }\sigma$ significance, making a broken power law model the preferred model in this case as well.
This result, when combined to the proton spectrum, suggests that the spectral features are more likely to be rigidity-depending rather than energy-depending, even though
the latter cannot be ruled out because of the uncertainties.

\begin{figure}[h]
\centering
\begin{subfigure}{.463\textwidth}
  \centering
  \includegraphics[width=\linewidth]{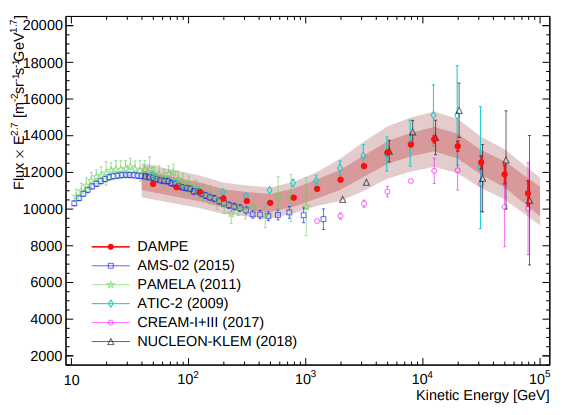}
  \caption{DAMPE Proton spectrum\cite{DAMPE_protons}}
  \label{fig:dampe_p}
\end{subfigure}
\begin{subfigure}{.48\textwidth}
  \centering
  \includegraphics[width=\linewidth]{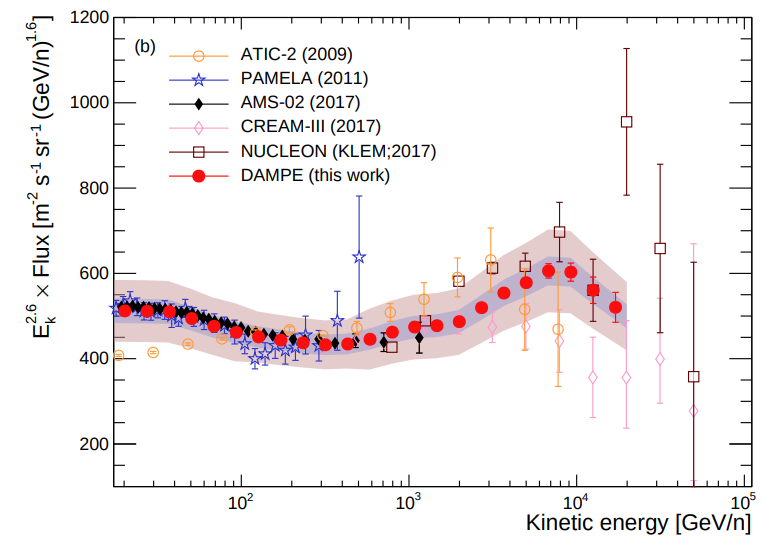}  
  \caption{DAMPE Helium spectrum\cite{DAMPE_helium}}
  \label{fig:dampe_He}
\end{subfigure}
\caption{DAMPE proton (a) and Helium (b) spectra, compared with measurements from various 
	experiments (\cite{AMS_protons}\cite{PAMELA_protons}\cite{ATIC_protons}\cite{CREAM_protons}\cite{NUCLEON_protons} 
	and \cite{PAMELA_protons}\cite{ATIC_protons}\cite{CREAM_protons}\cite{AMS_helium}\cite{NUCLEON_helium} respectively) }
\label{fig:dampe_p_He}
\end{figure}

\begin{figure}[h]
\centering
\includegraphics[width=0.5\textwidth]{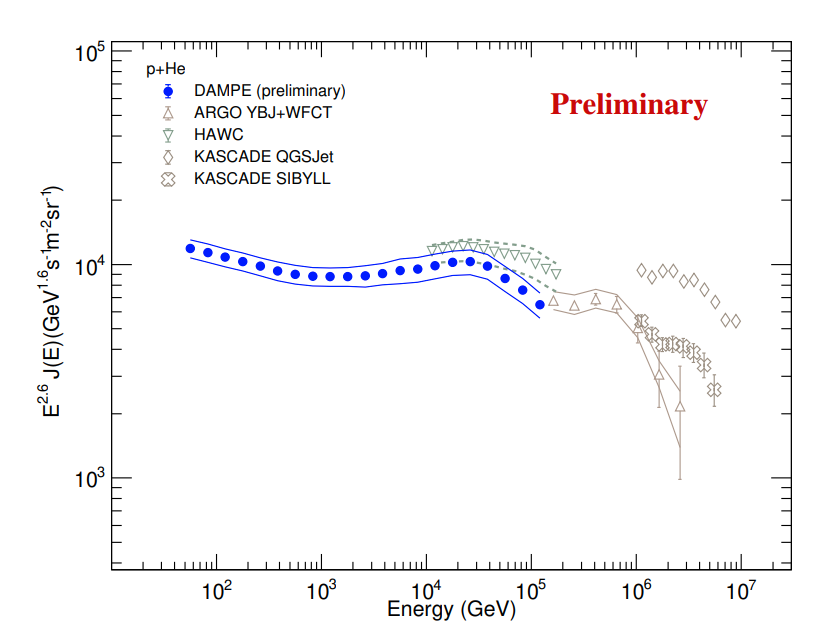}
\caption{DAMPE preliminary results of p+He spectrum\cite{DAMPE_pHe}, compared with HAWC\cite{HAWC_pHe} and more indirect 
	measurements\cite{ARGO_pHe}\cite{KASCADE_pHe}. The DAMPE energy range is approaching indirect experiments.}
\label{fig:dampe_pHe}
\end{figure}

%\paragraph{p+He}
In order to have more details about the behaviour of cosmic-ray spectrum at the highest direct measured energies, 
the combined analysis of protons and helium is being carried out\cite{DAMPE_pHe}. 
This analysis is characterised by a very high statistical sample with very low contamination, since the flux of the neighbouring nuclei is significantly lower, resulting in a high-energy spectrum that can also be compared with indirect measurements, both for the type of measurement and the energies. In particular,
the preliminary results shown in fig. \ref{fig:dampe_pHe} are already comparable to the ones provided by HAWC\cite{HAWC_pHe}, and getting close
to the energies of the first part ARGO-YBJ\cite{ARGO_pHe} flux. In the future, we might be able to reach an energy high enough to have our measurement almost 
comparable to KASCADE flux\cite{KASCADE_pHe} as well.
%, but from more recent advances on the analysis we are targeting 700 TeV, putting our last point after 
%the first points from ARGO-YBJ\cite{ARGO_pHe}, and going very close to KASCADE flux\cite{KASCADE_pHe}. 
Aside from the importance of the measurement per se,
this spectrum might be very useful as a
bridge between direct and indirect measurements, since
the former are not depending on uncertainties related to Extensive Air Shower: in particular the underlying hadronic interaction models.

%\begin{figure}[h]
%\centering
%\includegraphics[width=0.6\textwidth]{dampe_nuclei.png}
%\caption{The DAMPE charge resolution given by the PSD\cite{DAMPE_PSD} is accurate enough to distinguish 
%	most of the peaks related to the CR composition up to Nickel }
%\label{fig:dampe_nuclei}
%\end{figure}
%\includegraphics[]{dampe_b_over_c.png}

\begin{figure}[h]
\centering
\begin{subfigure}{0.40\textwidth}
  \centering
  \includegraphics[width=\linewidth]{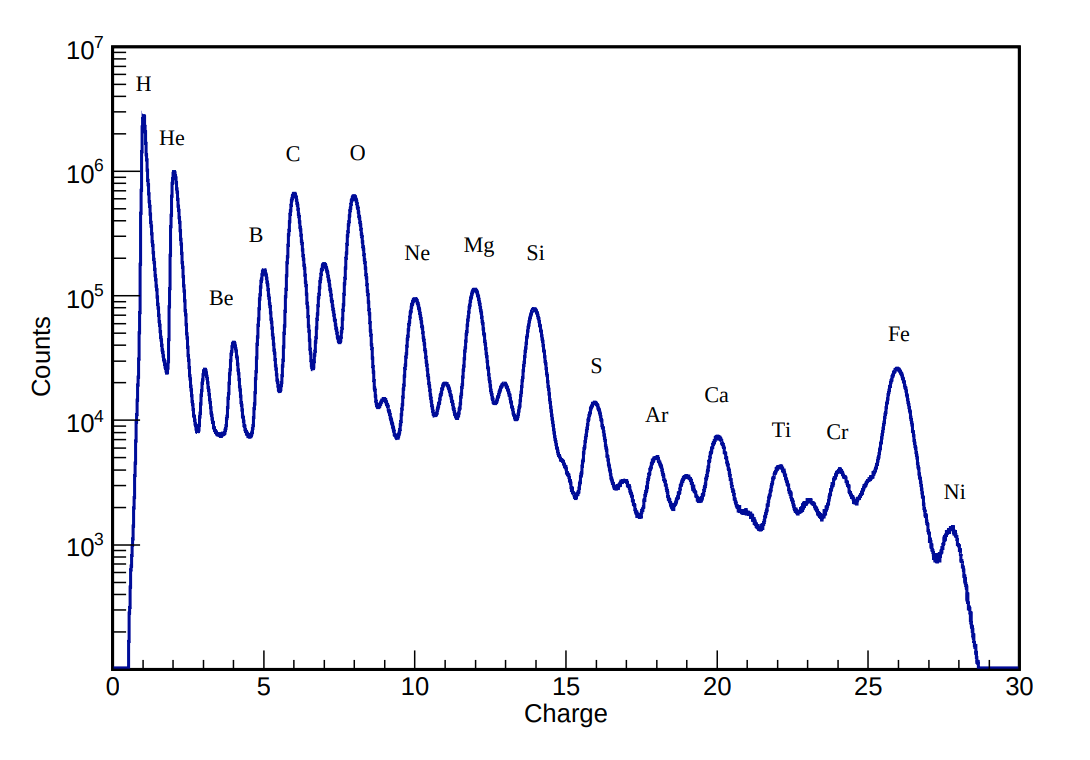}
  \caption{Charge reconstructed by PSD\cite{DAMPE_PSD}}
  \label{fig:dampe_nuclei_psd}
\end{subfigure}
\begin{subfigure}{.375\textwidth}
  \centering
  \includegraphics[width=\linewidth]{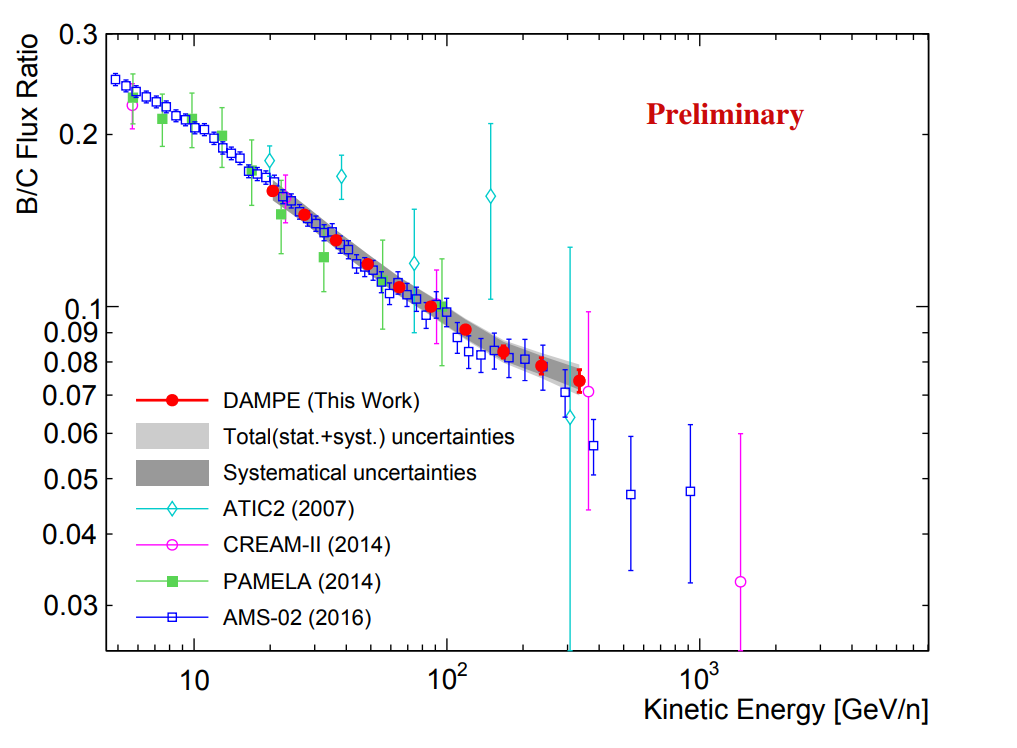}  
  \caption{Preliminary B/C flux ratio\cite{DAMPE_B_over_C}}
  \label{fig:dampe_nucle_b_o_c}
\end{subfigure}
\caption{The DAMPE charge resolution given by the PSD (a) is precise enough to distinguish 
	most of the peaks related to the CR composition up to Nickel. As an example, the preliminary Boron over Carbon flux ratio is shown (b)}
\label{fig:dampe_nuclei}
\end{figure}

\section{More analyses on heavier nuclei}
\label{sec:nuclei}
Several additional works on heavier nuclei are in progress with the aim of measuring their flux and the ratio among them. 
The nuclei are identified by the energy deposit
inside the PSD\cite{DAMPE_PSD}, with an example histogram of reconstructed charge from PSD shown in fig. \ref{fig:dampe_nuclei_psd}. 
Currently, the ongoing analyses include:
\begin{itemize}
	\item Single species spectra of light and intermediate mass nuclei, such as Boron and Carbon\cite{DAMPE_B_C};
	\item Spectra of heavy elements, like Fe\cite{DAMPE_Fe} and ultra-Fe species;
	\item Flux ratios, like the Boron over Carbon\cite{DAMPE_B_over_C} shown in fig. \ref{fig:dampe_nucle_b_o_c}.
\end{itemize}

\section{Conclusion}
The DAMPE mission has been continuously collecting data since December 2015, in stable conditions and with all its sub-detectors fully working. 
Data analysis produced already important results, like the proton\cite{DAMPE_protons} and helium\cite{DAMPE_helium} spectra shown in this work, which highlighted
new features in the cosmic-ray spectra, and the electrons\cite{DAMPE_electrons} and $\gamma$\cite{DAMPE_gamma} as well. Further analyses on other nuclei are currently being carried out.

\section*{Acknowledgements}
%\paragraph{Funding information}
The DAMPE mission was funded by Chinese Academy of Sciences (CAS). Activities are supported by National Key Research and Development Program of China (2016YFA0400200), National Natural Science Foundation of China (11921003, 11622327, 11722328, 11851305,
U1738205, U1738206, U1738207, U1738208, U1738127), strategic priority science and technology projects of CAS (XDA15051100), 100 Talents Program
of CAS, Young Elite Scientists Sponsorship Program by CAST
(YESS20160196), Program for Innovative Talents and Entrepreneur in Jiangsu, 
the Swiss National Science Foundation
(SNSF), National Institute for Nuclear Physics (INFN), Italy, and European Research Council (ERC) Horizon 2020 programme (851103).

%\paragraph{Author contributions}
%This work is the result of the contributions and efforts of all the participating institutes. All authors have reviewed, discussed, and commented on the results and on the manuscript. In line with the collaboration policy, the authors are listed here alphabetically.

% TODO: include funding information
%Authors are required to provide funding information, including relevant agencies and grant numbers with linked author's initials. Correctly-provided data will be linked to funders listed in the \href{https://www.crossref.org/services/funder-registry/}{\sf Fundref registry}.

\begin{appendix}

% TODO:
% Provide your bibliography here. You have two options:

% FIRST OPTION - write your entries here directly, following the example below, including Author(s), Title, Journal Ref. with year in parentheses at the end, followed by the DOI number.\right \texttt{•}

% \bibliographystyle{SciPost_bibstyle} % Include this style file here only if you are not using our template
%\bibliography{SciPost_Example_BiBTeX_File.bib}

\begin{thebibliography}{99}
%\bibitem{1931_Bethe_ZP_71} H. A. Bethe, {\it Zur Theorie der Metalle. i. Eigenwerte und Eigenfunktionen der linearen Atomkette}, Zeit. f{\"u}r Phys. {\bf 71}, 205 (1931), \doi{10.1007\%2FBF01341708}.
%\bibitem{arXiv:1108.2700} P. Ginsparg, {\it It was twenty years ago today... }, \url{http://arxiv.org/abs/1108.2700}.
\bibitem{DAMPE_mission} J. Chang et al., {\it The DArk Matter Particle Explorer mission}, Astroparticle Physics {\bf 95}, 205 (2017), 
	\doi{10.1016/j.astropartphys.2017.08.005}

	
% ---- PROTON SPECTRA ----- %

\bibitem{DAMPE_protons} Q. An et al., {\it Measurement of the cosmic ray proton spectrum from 40 GeV to 100 TeV with the DAMPE satellite}, Sci. Adv. {\bf 5}, 
eaax3793 (2019), \doi{10.1126/sciadv.aax3793}

\bibitem{AMS_protons} M. Aguilar et al., {\it Precision measurement of the proton flux in primary cosmic rays from rigidity 1 GV to 1.8 TV with the 
	alpha magnetic spectrometer on the international space station}, Phys. Rev. Lett. {\bf 114}, 171103 (2015), \doi{10.1103/PhysRevLett.114.171103}

\bibitem{PAMELA_protons} O. Adriani et al., {\it PAMELA measurements of cosmic-ray proton and helium spectra}, Science {\bf 332}, 69-72 (2011),
	\doi{10.1126/science.1199172}

\bibitem{ATIC_protons} A. D. Panov et al., {\it Energy spectra of abundant nuclei of primary cosmic rays from the data of ATIC-2 experiment: Final results},
	Bull. Russ. Acad. Sci. Phys. {\bf 73}, 564-567 (2009), \doi{10.3103/S1062873809050098}

\bibitem{CREAM_protons} Y. S. Yoon et al., {\it Proton and helium spectra from the CREAM-III flight}, Astrophys. J. {\bf 839}, 5 (2017),
	\doi{10.3847/1538-4357/aa68e4}

\bibitem{NUCLEON_protons} E. Atkin et al., {\it New universal cosmic-ray knee near a magnetic rigidity of 10 TV with the NUCLEON space observatory},
	JETP Lett. {\bf 108}, 5-12 (2018), \doi{10.1134/S0021364018130015}


% ---- HELIUM SPECTRA ----- %

\bibitem{DAMPE_helium} F. Alemanno et al., {\it Measurement of the Cosmic Ray Helium Energy Spectrum from 70 GeV to 80 TeV with the DAMPE Space Mission},
Phys. Rev. Lett. {\bf 126}, 201102 (2021), \doi{10.1103/PhysRevLett.126.201102}


\bibitem{AMS_helium} M. Aguilar et al., {\it Precision Measurement of the Helium Flux in Primary Cosmic Rays of Rigidities 1.9 GV to 3 TV with the Alpha Magnetic Spectrometer on the International Space Station}, Phys. Rev. Lett. {\bf 115}, 211101 (2015), \doi{10.1103/PhysRevLett.115.211101}


\bibitem{NUCLEON_helium} E. Atkin et al., {\it First results of the cosmic ray NUCLEON experiment}, J. Cosmol. Astropart. Phys. {\bf 07}, 020 (2017),
	\doi{10.1088/1475-7516/2017/07/020}


% ---- p+He SPECTRA ----- %

\bibitem{DAMPE_pHe} F. Alemanno et al., {\it Measurement of the light component (p+He) energy spectrum with the DAMPE space mission}, PoS ICRC2021 {\bf 395},
117 (2021), \doi{10.22323/1.395.0117}

\bibitem{HAWC_pHe} A. Albert et al., {\it Cosmic ray spectrum of protons plus helium nuclei between 6 and 158 TeV from HAWC data}, Phys. Rev. D {\bf 105}, 
063021 (2022), \doi{10.1103/PhysRevD.105.063021}

\bibitem{ARGO_pHe} B. Bartoli et al., {\it Knee of the cosmic hydrogen and helium spectrum below 1 PeV measured by ARGO-YBJ and a Cherenkov telescope of LHAASO},
Phys. Rev. D {\bf 92}, 092005 (2015), \doi{10.1103/PhysRevD.92.092005} 

\bibitem{KASCADE_pHe} K.-H. Kampert et al., {\it Cosmic rays in the 'knee'-region - Recent results from KASCADE}, Acta Phys.Polon. B {\bf 35}, 1799-1812 (2004)


% ---- nuclei ----- %

\bibitem{DAMPE_PSD} T. Dong et al., {\it Charge measurement of cosmic ray nuclei with the plastic scintillator detector of DAMPE}, Astroparticle Physics {\bf 105}, 
31 (2019), \doi{10.1016/j.astropartphys.2018.10.001}

\bibitem{DAMPE_B_C} W. Libo et al., {\it Towards the measurement of carbon and oxygen spectra in cosmic rays with DAMPE}, PoS ICRC2021 {\bf 395}, 128 (2021),
\doi{10.22323/1.395.0128}

\bibitem{DAMPE_Fe} Z. Xu et al., {\it Direct Measurement of the Cosmic-Ray Iron Spectrum with the Dark Matter Particle Explorer}, PoS ICRC2021 {\bf 395}, 
126 (2021), \doi{10.22323/1.395.0115}

\bibitem{DAMPE_B_over_C} Z.F. Chen et al, {\it Measurement of the Boron to Carbon Flux Ratio in Cosmic Rays with the DAMPE Experiment}, PoS ICRC2021 {\bf 395}, 
115 (2021), \doi{10.22323/1.395.0126}

\bibitem{DAMPE_electrons} G. Ambrosi et al, {\it Direct detection of a break in the teraelectronvolt cosmic-ray spectrum of electrons and positrons}, Nature {\bf 552}, 63–66 (2017), \doi{10.1038/nature24475}

\bibitem{DAMPE_gamma} F. Alemanno et al, {\it Search for gamma-ray spectral lines with the DArk Matter Particle Explorer}, Science Bulletin {\bf 67}, 7 (2022), \doi{10.1016/j.scib.2021.12.015}

\end{thebibliography}

\nolinenumbers
\end{appendix}

\end{document}